\documentclass[fleqn,12pt,twoside]{article}
\usepackage[headings]{espcrc1}

\readRCS
$Id: espcrc1.tex,v 1.2 2004/02/24 11:22:11 spepping Exp $
\usepackage{graphicx}
\usepackage[figuresright]{rotating}

\newcommand{\AmS}{{\protect\the\textfont2
  A\kern-.1667em\lower.5ex\hbox{M}\kern-.125emS}}

\title{On Double-Beta Decay Half-Life Time Systematics}

\author{B.\ Pritychenko\address[BNLab]{National Nuclear Data Center, Brookhaven National Laboratory, Upton, NY 11973-5000, U.S.A.}
}
       
\runtitle{On Double-Beta Decay Half-Life Time Systematics}
\runauthor{B. Pritychenko}

\begin{document}

% typeset front matter
\maketitle

\begin{abstract}
Recommended $2 \beta$(2$\nu$) half-life values and their systematics were analyzed in the framework of a simple empirical approach. 
\mbox{$T^{2 \nu}_{1/2}$ $\sim$ $\frac{1}{E^{8}}$} trend has been observed for $^{128,130}$Te recommended values. 
This trend was used to predict $T^{2 \nu}_{1/2}$ for all isotopes of interest. 
Current results were compared with other theoretical and experimental works. 
\end{abstract}

\section{Introduction}

Double-beta decay was originally proposed by M. Goeppert-Mayer in 1935 \cite{35Go} as a nuclear disintegration with simultaneous emission of  two electrons and two neutrinos: 
\begin{equation}
\label{myeq.2b}  
(Z,A) \rightarrow (Z+2,A) + 2 e^{-} + ( 0\  or \  2)\bar{\nu}_{e}
\end{equation}

 There are four possible double-beta decay processes: $2 \beta^{-}$,  $2 \beta^{+}$, $\epsilon$ $\beta^{+}$, 2$\epsilon$ and two decay modes: 
 two-neutrino (2$\nu$) and neutrinoless (0$\nu$)  \cite{87Bo,02El,05Su}. 
 2$\nu$-mode is not prohibited by any conservation law and definitely occurs as a second-order process compared to the regular $\beta$-decay. 0$\nu$-mode differs from the 2$\nu$-mode by the fact that no neutrinos are emitted during the decay. This normally requires that lepton number is not conserved and neutrino should contain a small fraction of massive particles that equals to its anti-particles (Majorana neutrino). Obviously, observation of $2 \beta$(0$\nu$)-decay will have enormous implications on particle physics and fundamental symmetries. While observation of $2 \beta$(2$\nu$)-decay will provide information on nuclear structure physics that can be used in 0$\nu$-mode calculations \cite{05Ci}.  

Experimental evidence and theoretical calculations indicate that probability for 2$\nu$-mode is much higher than for 0$\nu$-mode. 
In fact, $^{76}$Ge $2 \beta$-decay experiments  \cite{90Va,90Mi,02Aa,01Kl,05Ba} have demonstrated that half-life time for $2 \beta$(2$\nu$)-decay is at least four 
orders of magnitude lower than $2 \beta$(0$\nu$). 
Finally, we concentrate here on the experimentally observed 2$\nu$-mode only.

\section{Analysis of Recommended Values}

Double-beta decay is an important physical process and experimental results, in this field, have been compiled by several groups \cite{02Tr,nndc,pdg}. 
These compilations were used to produce adopted or recommended values \cite{06Ba,08Pr}. 
Table \ref{table1} shows  NNDC-recommended values \cite{08Pr}  which were deduced in the accordance with the U.S. Nuclear Data Program guidelines \cite{Lwei}. 
NNDC recommended numbers represent the best available values, further measurements will result in the addition of new and improvements to existing values.

\begin{table}[htbp]
\centering
\caption{NNDC-recommended $2 \beta$-decay values. Data are taken from \cite{nndc,08Pr,01Ra,03Au}.}
\begin{tabular}{c|c|c|c|c|c|c}
\hline
\hline
Nuclide  &  Process & Transition & E (keV) & $\beta_{2}$ & T$_{1/2}^{2 \nu}$(y) &  T$_{1/2}^{0 \nu + 2 \nu}$(y)  \\
\hline
\hline

$^{48}$Ca & $2 \beta^{-}$ & 0$^{+}$ $\rightarrow$ 0$^{+}$ & 4273.6 & 0.106 & (4.3$\pm$2.3)x10$^{19}$ & \\ 		
$^{76}$Ge & $2 \beta^{-}$  & 0$^{+}$ $\rightarrow$ 0$^{+}$ & 2039.0 & 0.2623 & (1.3$\pm$0.4)x10$^{21}$ &  \\ 		
$^{82}$Se & $2 \beta^{-}$  & 0$^{+}$ $\rightarrow$ 0$^{+}$ &  2995.5 & 0.1934 &(9.2$\pm$0.8)x10$^{19}$ & \\ 		
$^{96}$Zr & $2 \beta^{-}$  & 0$^{+}$ $\rightarrow$ 0$^{+}$ & 3347.7 & 0.080 & (2.0$\pm$0.4)x10$^{19}$ &  \\ 		
$^{100}$Mo & $2 \beta^{-}$  & 0$^{+}$ $\rightarrow$ 0$^{+}$ &  3034.68 & 0.2309 & (7.0$\pm$0.4)x10$^{18}$ &  \\ 		
$^{100}$Mo & $2 \beta^{-}$  & 0$^{+}$ $\rightarrow$ 0$^{+}_{1}$ & 2339.6 & 0.2309 & (5.7$\pm$1.4)x10$^{20}$ & \\ 		
$^{100}$Mo & $2 \beta^{-}$  & 0$^{+}$ $\rightarrow$ 0$^{+}_{1}$ & 2339.6 & 0.2309 & & (6.1$\pm$0.2)x10$^{20}$ \\
$^{116}$Cd & $2 \beta^{-}$  & 0$^{+}$ $\rightarrow$ 0$^{+}$ & 2808.7 & 0.1906 & (3.0$\pm$0.3)x10$^{19}$ &  \\ 		
$^{128}$Te & $2 \beta^{-}$  & 0$^{+}$ $\rightarrow$ 0$^{+}$ & 867.95 & 0.1363 & & (3.5$\pm$2.0)x10$^{24}$ \\
$^{130}$Te & $2 \beta^{-}$  & 0$^{+}$ $\rightarrow$ 0$^{+}$ & 2530.3 & 0.1184 & (6.1$\pm$4.8)x10$^{20}$ &  \\ 		
$^{130}$Ba & 2$\epsilon$ & 0$^{+}$ $\rightarrow$ 0$^{+}$ & 2620.1 & 0.2183 &  & (2.2$\pm$0.5)x10$^{21}$ \\
$^{150}$Nd & $2 \beta^{-}$  & 0$^{+}$ $\rightarrow$ 0$^{+}$ &  3367.68 & 0.2853 & (7.9$\pm$0.7)x10$^{18}$ & \\ 		
$^{150}$Nd & $2 \beta^{-}$  & 0$^{+}$ $\rightarrow$ 0$^{+}_{1}$ &  2692.3 & 0.2853 &  & (1.4$\pm$0.5)x10$^{20}$ \\
$^{238}$U  & $2 \beta^{-}$  & 0$^{+}$ $\rightarrow$ 0$^{+}$ & 1144.2 & 0.2863 & & (2.0$\pm$0.6)x10$^{21}$ \\
\hline
\hline
\end{tabular}
\label{table1}
\end{table} 

$T_{1/2}^{2\nu}$ values are often described as follows \cite{87Bo}:
\begin{equation}
\label{myeq.Half}  
(T_{1/2}^{2\nu} (0^{+} \rightarrow 0^{+}))^{-1} = G^{2 \nu} (E,Z) \times |M^{2 \nu}_{GT} - \frac{g^{2}_{V}}{g^{2}_{A}} M^{2 \nu}_{F}|^{2}, 
\end{equation}
where the function  $G^{2 \nu} (E,Z)$ results from lepton phase space integration and contains all relevant constants. 
From the Eq. \ref{myeq.Half} one may conclude that decay half-lives depend on transition energy, charge and nuclear deformation.

It will be useful to analyze these half-lives using the Grodzins' approach \cite{62Gr} and the relevant data from Table \ref{table1}. 
In this analysis, we will consider only  $2 \beta^{-}$-decay parameters for   0$^{+}$ $\rightarrow$ 0$^{+}$ transitions, i.e. transitions 
without $\gamma$-rays and adopt deformation parameters ($\beta_{2}$) from Raman {\it et al.} \cite{01Ra}. 
Table \ref{table1} indicates that in spite of small data sample we are effectively covering the whole range of nuclei from Z=20 to Z=92.

%\begin{table}
%\centering
%\caption{$\beta^{-}$$\beta^{-}$-decay parameters for  0$^{+}$ $\rightarrow$ 0$^{+}$ transitions. Data are taken from \cite{08Pr,01Ra,03Au}.}
%\begin{tabular}{c|c|c|c|c|c|c}
%\hline
%\hline
% & Nuclide  &  Z & Transition Energy (keV) & $\beta_{2}$ & T$_{1/2}^{2 \nu}$(y) &  T$_{1/2}^{0 \nu + 2 \nu}$(y)  \\
%\hline
%\hline
%
%1 &     $^{48}$Ca & 20 & 4273.60$\pm$4.00 & 0.106(18) & (4.3$\pm$2.3)x10$^{19}$  & \\ 		
%2 &	$^{76}$Ge & 32 & 2039.00$\pm$0.05 & 0.2623(39) & (1.3$\pm$0.4)x10$^{21}$ & \\ 		
%3 &	$^{82}$Se & 34 & 2995.50$\pm$1.87 & 0.1934(27) & (9.2$\pm$0.8)x10$^{19}$  & \\ 		
%4 & 	$^{96}$Zr & 40 & 3347.70$\pm$2.20 & 0.080(17) & (2.0$\pm$0.4)x10$^{19}$  & \\ 		
%5 &	$^{100}$Mo & 42 & 3034.68$\pm$5.91 & 0.2309(22) & (7.0$\pm$0.4)x10$^{18}$  & \\ 		
%6 &	$^{116}$Cd & 48 & 2808.71$\pm$3.72 & 0.1906(34) &(3.0$\pm$0.3)x10$^{19}$  & \\ 		
%7 &	$^{128}$Te & 52 & 867.95$\pm$1.47 & 0.1363(11) & & (3.5$\pm$2.0)x10$^{24}$ \\
%8 &	$^{130}$Te & 52 & 2530.30$\pm$1.99 & 0.1184(14) & (6.1$\pm$4.8)x10$^{20}$  & \\ 		
%9 &	$^{150}$Nd & 60 & 3367.68$\pm$2.21 & 0.2853(21) & (7.9$\pm$0.7)x10$^{18}$  & \\ 		
%10 &	$^{238}$U  & 92 & 1144.20$\pm$1.22 & 0.2863(24) & & (2.0$\pm$0.6)x10$^{21}$ \\
%\hline
%\hline
%\end{tabular}
%\label{table2}
%\end{table}

First, we will analyze half-life values of $^{128,130}$Te. Both tellurium isotopes have the same charge, similar nuclear structure while $2 \beta^{-}$-transition energies are different. 
It is natural to assume that difference between tellurium half-lives is due to transition energies \cite{68Po}. In fact, tellurium data for $T_{1/2}^{2\nu}$ are consistent with the following ratio:
\begin{equation}
\label{myeq.Tel0}  
T_{1/2}^{2\nu} (^{128}Te) / T_{1/2}^{2\nu} (^{130}Te) \approx 5.7 \times 10^{3} \sim (\frac{E_{130}}{E_{128}})^{8.1}
\end{equation}
From here we deduce the following trend: 
\begin{equation}
\label{myeq.Tel}  
T_{1/2}^{2\nu} (0^{+} \rightarrow 0^{+}) \sim \frac{1}{E^8}
\end{equation} 

This conclusion agrees  well with the theoretical calculation of Primakoff and Rosen \cite{69Pr} who predicted that for $\beta$$\beta$(2$\nu$) decay, the phase space available to the (four) emitted leptons is roughly proportional to the eighth through 11$^{th}$ power of energy release. 

It is known that half-lives depend on dimensionless Coulomb energy parameter \mbox{$\xi$ $\approx$ $Z A^{-1/3}$} \cite{98Bo}. 
%\begin{equation}
%\label{myeq.Coul}  
%\xi \approx Z A^{-1/3}
%\end{equation}
This will modify our result for $2 \beta$-transition as follows:
\begin{equation}
\label{myeq.Tel2}  
T_{1/2}^{2\nu} (0^{+} \rightarrow 0^{+}) \sim \frac{1}{E^8 \xi^2}
\end{equation} 

Second, we notice that half-life value for $^{100}$Mo is lower than for $^{96}$Zr while deformation parameter ($\beta_{2}$) for $^{100}$Mo is almost 3 times larger than that of $^{96}$Zr. In this work, we will try to model dependence of halflives on nuclear deformation with deformation parameters ($\beta_{2}$). These two half-lives  become more consistent if we will  include deformation of the parent nucleus into Eq. \ref{myeq.Tel2}:
\begin{equation}
\label{myeq.Zr}  
T_{1/2}^{2\nu} (0^{+} \rightarrow 0^{+}) \sim \frac{1}{E^{8} \xi^{2} \beta_{2}^{2}}
\end{equation} 

\begin{figure}
\begin{center}
\includegraphics[height=7cm]{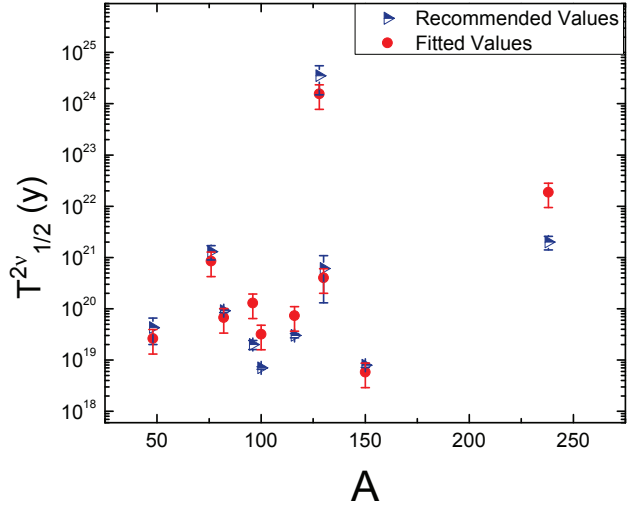}
\caption{NNDC recommended and fitted T$^{2 \nu}_{1/2}$ values.}
\label{fit0}
\end{center}
\end{figure}

Finally, a fit of experimental data provides the empirical rule for T$_{1/2}^{2\nu}$:
\begin{equation}
\label{myeq.Fit}  
T_{1/2}^{2\nu} (0^{+} \rightarrow 0^{+}) \approx \frac{(1\pm0.5)\times 10^{24}}{E^{8} \xi^{2} \beta_{2}^{2}},
\end{equation}
where $T_{1/2}^{2\nu}$ in years and $E$ in MeV. Large error in the Eq. \ref{myeq.Fit} is due to deformation parameter values, where $\sim$20 \% of $\beta_{2}$ error will result in $\sim$40 \% deviation of  $T_{1/2}^{2\nu}$. 
Fit results are shown in Fig. \ref{fit0}.

%There is a reasonably good agreement between recommended and predicted values with a possible exception of  $^{96}$Zr, $^{100}$Mo, $^{238}$U and perhaps $^{116}$Cd. This may indicate current model applicability issues for heavy nuclei or problems with the experiment.  %$^{96}$Zr and $^{238}$U results are based on single  measurements \cite{99Ar,91Tu}, additional measurements may improve this situation. 

Empirical rule, Eq. \ref{myeq.Fit} is only valid for $2 \beta$(2$\nu$)-decay transitions without $\gamma$-rays. For transitions to the excited 0$^{+}$ states of daughter nuclei, when reaction products  are emitted and statistical sample is reduced to $^{100}$Mo and $^{150}$Nd, half-life will depend on higher than eight power of energy release. 
%\mbox{$T^{2 \nu}_{1/2}$ $\sim$ $\frac{1}{E^{11}}$}.

\section{Empirical Rule Predictions}
Eq. \ref{myeq.Fit} allows us to predict half-life times for all nuclei of interest as shown in Table \ref{table2}.  In most cases, they agree reasonably well with recommended and 
experimental values with exception of $^{136}$Xe,  $^{238}$U, $^{96}$Zr, $^{100}$Mo and perhaps $^{116}$Cd. 

There is an acute problem with $^{136}$Xe  that is far more stable than comparable nucleus of $^{130}$Te. 
%Neutron shell closure at $N$=82 in $^{136}$Xe points explains the small degree of deformation.  
Under current assumptions, $^{136}$Xe calculated half-life is very sensitive to the deformation parameter ($\beta_{2}$) that was adopted from Raman {\it et al.} \cite{01Ra}. 
Unfortunately, due to lack of experimental data, Raman's values are often based on a single measurement \cite{01Ra,be2}. For example,  $\beta_{2}$($^{136}$Xe) was deduced from the measurement of Speidel \cite{93Sp}. By replacing Raman's adopted value with an earlier measurement of Edvardson and Norlin \cite{75Ed} we obtain  T$_{1/2}^{2 \nu, predicted}$ $\approx$ (9.0$\pm$4.5)$\times$10$^{20}$ y that is comparable with the experiment \cite{02Be,06Ga}. The similar situation is observed for  $^{96}$Zr, $^{100}$Mo, $^{238}$U and $^{116}$Cd, where Raman values are based on the relatively old and sometimes inconsistent data \cite{01Ra}. 

Current analysis highlights  strong dependence of $2 \beta$(2$\nu$)-decay half-life values on nuclear deformation. Future measurements with charge-exchange reactions will help to clarify these issues. This will lead to better understanding of nuclear structure of even-even nuclei. The same analysis also reveals high probability for $2 \beta$($2 \nu$)-decay in the strongly deformed nucleus of $^{160}$Gd. 
\begin{table}
\centering
\caption{Empirical Rule Predictions: $2 \beta^{-}$(2$\nu$)-decay predicted, recommended and experimental values for  0$^{+}$ $\rightarrow$ 0$^{+}$ transitions. $0 \nu$ contribution was ignored for two cases (*). Data are taken from \cite{01Ra,03Au,08Pr,nndc}.}
\begin{tabular}{c|c|c|c|c|c}
\hline
\hline
  Nuclide  &  E (keV) & $\beta_{2}$ & T$_{1/2}^{2 \nu, predicted}$(y) &  T$_{1/2}^{2 \nu, recommended}$(y)  &  T$_{1/2}^{2 \nu , experimental}$(y)  \\
\hline
\hline
$^{46}$Ca &	988.35 &	0.153 &		(1.50$\pm$0.75)x$10^{24}$ & & \\
$^{48}$Ca &	4273.6 &	0.106 &		(2.63$\pm$1.32)x$10^{19}$ &  (4.3$\pm$2.3)x10$^{19}$ & 4.3x10$^{19}$ \\
$^{70}$Zn &	998.46 &	0.228 &		(3.67$\pm$1.83)x$10^{23}$ & & $>$1.3$\times$10$^{16}$\\
$^{76}$Ge &	2039.0 &	0.2623 &	(8.50$\pm$4.25)x$10^{20}$ & (1.3$\pm$0.4)x10$^{21}$ & 1.55$\times$10$^{21}$ \\
$^{80}$Se &	132.56 &	0.2318 &	(3.13$\pm$1.56)x$10^{30}$ & & \\
$^{82}$Se &	2995.5 &	0.1934 &	(6.71$\pm$3.36)x$10^{19}$ & (9.2$\pm$0.8)x10$^{19}$ & 9.6$\times$10$^{19}$ \\
$^{86}$Kr &	1258.01 &	0.145 &		(1.14$\pm$0.57)x$10^{23}$ & & \\
$^{94}$Zr &	1142.87 &	0.09 &		(5.46$\pm$2.73)x$10^{23}$ & & $>$1.9$\times$10$^{19}$  \\
$^{96}$Zr &	3347.7 &	0.08 &		(1.29$\pm$0.65)x$10^{20}$ & (2.0$\pm$0.4)x10$^{19}$  & 2.0$\times$10$^{19}$ \\
$^{98}$Mo &	112.75 &	0.1683 &	(1.62$\pm$0.81)x$10^{31}$ & & \\
$^{100}$Mo &	3034.68 &	0.2309 &	(3.18$\pm$1.59)x$10^{19}$ & (7.0$\pm$0.4)x10$^{18}$ & 7.11$\times$10$^{18}$ \\
$^{104}$Ru &	1301.17 &	0.2707 &	(1.89$\pm$0.95)x$10^{22}$ & & \\
$^{110}$Pd &	2003.8 &	0.257 &		(6.30$\pm$3.15)x$10^{20}$ & & $>$6.0$\times$10$^{16}$ \\
$^{114}$Cd &	539.96 &	0.1903 &	(3.89$\pm$1.94)x$10^{25}$ & & $\geq$1.3$\times$10$^{18}$ \\
$^{116}$Cd &	2808.7 &	0.1906 &	(7.31$\pm$3.66)x$10^{19}$ &  (3.0$\pm$0.3)x10$^{19}$ & 2.8$\times$10$^{19}$ \\
$^{122}$Sn &	368.08 &	0.1036 &	(2.71$\pm$1.36)x$10^{27}$ & & \\
$^{124}$Sn &	2287.81 &	0.0953 &	(1.45$\pm$0.73)x$10^{21}$ & & $>$1.0$\times$10$^{17}$ \\
$^{128}$Te &	867.95 &	0.1363 &	(1.56$\pm$0.78)x$10^{24}$ & (3.5$\pm$2.0)x10$^{24}$* & 2.2$\times$10$^{24}$* \\
$^{130}$Te &	2530.3 &	0.1184 &	(4.02$\pm$2.01)x$10^{20}$ & (6.1$\pm$4.8)x10$^{20}$ & 6.1$\times$10$^{20}$ \\
$^{134}$Xe &	825.38 &	0.119 &		(2.93$\pm$1.47)x$10^{24}$ & & $>$1.1$\times$10$^{16}$ \\
$^{136}$Xe &	2461.8 &	0.122 &		(4.50$\pm$2.25)x$10^{20}$ & & $\geq$8.5$\times$10$^{21}$ \\
$^{142}$Ce &	1416.72 &	0.1277 &	(3.05$\pm$1.52)x$10^{22}$ & & $>$1.6$\times$10$^{17}$ \\
$^{146}$Nd &	70.83 &		0.1524 &	(5.22$\pm$2.61)x$10^{32}$ & & \\
$^{148}$Nd &	1928.77 &	0.2013 &	(9.98$\pm$4.99)x$10^{20}$ & & \\
$^{150}$Nd &	3367.68 &	0.2853 &	(5.80$\pm$2.90)x$10^{18}$ & (7.9$\pm$0.7)x10$^{18}$ & 7.7$\times$10$^{18}$ \\
$^{154}$Sm &	1251.62 &	0.341 &		(1.06$\pm$0.53)x$10^{22}$ & & \\
$^{160}$Gd &	1729.44 &	0.3534 &	(7.17$\pm$3.59)x$10^{20}$ & & $>$1.9$\times$10$^{19}$ \\
$^{170}$Er &	654.35 &	0.3363 &	(1.74$\pm$0.87)x$10^{24}$ & & \\
$^{176}$Yb &	1083.38 &	0.305 &		(3.62$\pm$1.81)x$10^{22}$ & & \\
$^{186}$W &	489.94 &	0.2257 &	(3.51$\pm$1.75)x$10^{25}$ & & $\geq$3.7$\times$10$^{18}$ \\
$^{192}$Os &	412.36 &	0.1667 &	(2.47$\pm$1.24)x$10^{26}$ & & \\
$^{198}$Pt &	1046.77 &	0.1141 &	(2.96$\pm$1.48)x$10^{23}$ & & \\
$^{204}$Hg &	419.49 &	0.0686 &	(1.20$\pm$0.60)x$10^{27}$ & & \\
$^{232}$Th &	837.57 &	0.2608 &	(2.82$\pm$1.41)x$10^{23}$ & & $>$2.1$\times$10$^{9}$ \\
$^{238}$U &	1144.2 &	0.2863 &	(1.88$\pm$0.94)x$10^{22}$ & (2.0$\pm$0.6)x10$^{21}$* & 2.0$\times$10$^{21}$* \\
\hline
\hline
\end{tabular}
\label{table2}
\end{table}

\section{Conclusion}
$^{128,130}$Te data analysis led to observation of $\frac{1}{E^{8}}$ energy trend for  T$_{1/2}^{2 \nu}$ recommended values that is consistent with two-nucleon mechanism of $2 \beta$(2$\nu$)-decay \cite{83Ab}. 
The energy trend and deformation parameters were used to explain nuclear systematics of recommended values 
and create an empirical rule for 2$\beta$(2$\nu$)-decay half-lives. The rule has been used to calculate  T$_{1/2}^{2 \nu}$ for all nuclei of interest. 

\section{Acknowledgments}                               
The author is grateful to A. Volya and M. Blennau for  productive discussions and  careful reading of the manuscript and useful suggestions, respectively. This work was funded by the Office of Nuclear Physics, Office of Science of the U.S. Department of Energy, under Contract No. DE-AC02-98CH10886 with Brookhaven Science Associates, LLC.  % here.


\newpage
\begin{thebibliography}{00}

\bibitem{35Go} M. Goeppert-Mayer, Phys. Rev. {\bf 48}, 512 (1935).
\bibitem{87Bo} F. Boehm and P. Vogel, Physics of Massive Neutrinos, Cambridge University Press, Cambridge, England (1987). 
\bibitem{02El} S.R. Elliott, P. Vogel, Ann. Rev. Nucl. Part. Sci. {\bf 52}, 115 (2002).
\bibitem{05Su} J. Suhonen, Nucl. Phys. {\bf A 752}, 53c (2005).
\bibitem{05Ci} O. Civitarese, J. Suhonen, Nucl. Phys. {\bf A 761}, 313 (2005). 
\bibitem{90Va} A.A. Vasenko, I.V. Kirpichnikov, V.A. Kuznetsov {\it et al.}, Mod. Phys. Lett. {\bf A5}, 1299 (1990).
\bibitem{90Mi} H.S. Miley, F.T. Avignone III, R.L. Brodzinski {\it et al.}, Phys. Rev. Lett. {\bf 65}, 3092 (1990).
\bibitem{02Aa} C.E. Aalseth, F.T. Avignone III, R.L. Brodzinski {\it et al.}, Phys. Rev. {\bf D65}, 092007 (2002).
\bibitem{01Kl} H.V. Klapdor-Kleingrothaus, A. Dietz, L. Baudis {\it et al.}, Eur. Phys. J. {\bf A 12}, 147 (2001).
\bibitem{05Ba} A.M. Bakalyarov, A.Ya. Balysh, S.T. Belyaev {\it et al.}, Part. and Nucl., Lett. {\bf 125}, 21 (2005). 
\bibitem{02Tr} V.I. Tretyak, Y.G. Zdesenko, At. Data Nucl. Data Tables {\bf 80}, 83 (2002).
\bibitem{nndc} NNDC $2 \beta$-decay Data Project, Available from $\langle$http://www.nndc.bnl.gov/bbdecay$\rangle$.
\bibitem{pdg}  C. Amsler, M. Doser, M. Antonelli {\it et al.}, Phys. Lett. {\bf B 667}, 1 (2008).
\bibitem{06Ba} A.S. Barabash, Czech. J. Phys. {\bf 56}, 437 (2006).
\bibitem{08Pr} B. Pritychenko, Nuclear Structure 2008, June 3-6, 2008, East Lansing, MI.
\bibitem{Lwei} E. Browne, Limitation of Relative Statistical Weight  Method, INDC (NDS) {\bf 363}, IAEA,  Vienna (1998).
\bibitem{62Gr} L. Grodzins, Phys. Lett. {\bf 2}, 88 (1962).
\bibitem{01Ra} S. Raman, C.W. Nestor, Jr., P. Tikkanen, At. Data Nucl. Data Tables {\bf 78}, 1 (2001). 
\bibitem{03Au} G. Audi, A.H. Wapstra, C. Thibault, Nucl. Phys. {\bf A729}, 337 (2003). 
\bibitem{68Po} B. Pontecorvo, Phys. Lett {\bf B 26}, 630 (1968).
\bibitem{69Pr} H. Primakoff and S.P. Rosen, Phys. Rev. {\bf 184}, 1925 (1969).
\bibitem{98Bo} A. Bohr, B.R. Mottelson, Nuclear Structure {\bf I}, World Scientific, Singapore (1998).
%\bibitem{99Ar} R. Arnold, C. Augier, J. Baker {\it et al.}, Nucl. Phys. {\bf A 658}, 299 (1999).
%\bibitem{91Tu} A.L. Turkevich, T.E. Economou, G.A. Cowan, Phys. Rev. Lett. {\bf 67}, 3211 (1991).
\bibitem{be2}  NNDC B(E2) Project, Available from $\langle$http://www.nndc.bnl.gov/be2$\rangle$.
\bibitem{93Sp} K.-H. Speidel, H. Busch, S. Kremeyer {\it et al.}, Nucl. Phys. {\bf A 552}, 140 (1993).
\bibitem{75Ed} L.-O. Edvardson, L.O. Norlin, Priv. Comm. (March 1975); UUIP-895 (1975).
\bibitem{02Be} R. Bernabei, P. Belli, F. Cappella {\it et al.}, Phys.Lett. {\bf 546B}, 23 (2002).
\bibitem{06Ga} Ju.M. Gavriljuk, A.M. Gangapshev, V.V. Kuzminov {\it et al.}, Phys. Atomic Nuclei {\bf 69}, 2129 (2006).
\bibitem{83Ab} J. Abad, A. Morales, R. Nunez-Lagos, A.F. Pacheco, Il Nuevo Cimento {\bf 75A}, 173 (1983).
\end{thebibliography}
\end{document}